\documentclass[prd,aps,a4,onecolumn,superscriptaddress,preprintnumbers,nofootinbib,notitlepage]{revtex4-1}

\usepackage{color}
\usepackage{cancel}
\usepackage{slashed}
\usepackage{amssymb}
\usepackage{amsmath}
\usepackage{hyperref}
\usepackage{enumerate}
\usepackage{multirow}
\usepackage{ulem}
\usepackage{float}
\usepackage{comment}
\usepackage{diagbox}
\usepackage{amsmath}
\usepackage[capitalize]{cleveref}

\usepackage{multirow}
\usepackage{tabularx}
\usepackage{graphicx}
\usepackage{subfigure}
\usepackage{caption}
\usepackage{comment}

\begin{document}

\preprint{MI-HET-836}

\title{\texorpdfstring{$\nu_\mu$ and $\nu_\tau$}{Mu- and tau-neutrino} elastic scattering in Borexino}

\author{Kevin J. Kelly}
\email{kjkelly@tamu.edu}
\affiliation{Department of Physics and Astronomy, Mitchell Institute for Fundamental Physics and Astronomy, Texas A\&M University, College Station, Texas 77843, USA}
\author{Nityasa Mishra}
\email{nityasa\_mishra@tamu.edu}%
\affiliation{Department of Physics and Astronomy, Mitchell Institute for Fundamental Physics and Astronomy, Texas A\&M University, College Station, Texas 77843, USA}
\author{Mudit Rai}
\email{muditrai@tamu.edu}%
\affiliation{Department of Physics and Astronomy, Mitchell Institute for Fundamental Physics and Astronomy, Texas A\&M University, College Station, Texas 77843, USA}
\author{Louis E. Strigari}%
\email{strigari@tamu.edu}%
\affiliation{Department of Physics and Astronomy, Mitchell Institute for Fundamental Physics and Astronomy, Texas A\&M University, College Station, Texas 77843, USA}

\date{\today}

\begin{abstract}
We perform a detailed study of neutrino-electron elastic scattering using the mono-energetic $^{7}$Be neutrinos in Borexino, with an emphasis on exploring the differences between the contributions of $\nu_e$, $\nu_\mu$, and $\nu_\tau$. We find that current data are capable of measuring these components such that the contributions from $\nu_\mu$ and $\nu_\tau$ cannot be zero, although distinguishing between them is challenging -- the differences stemming from Standard Model radiative corrections are insufficient without significantly more precise measurements. In studying these components, we compare predicted neutrino-electron scattering event rates within the Standard Model (accounting for neutrino oscillations), as well as going beyond the Standard Model in two ways. We allow for non-unitary evolution to modify neutrino oscillations, and find that with a larger exposure (${\sim}30$x), Borexino may provide relevant information for constraining non-unitarity, and that JUNO may be able to accomplish this with its data collection of $^{7}$Be neutrinos. We also consider novel $\nu_\mu$- and $\nu_\tau$-electron scattering from a gauged $U(1)_{L_\mu - L_\tau}$ model, showing consistency with previous analyses of Borexino and this scenario, but also demonstrating the impact of uncertainties on Standard Model mixing parameters on these results.
\end{abstract}

\maketitle

\section{Introduction \label{sec:introduction}}
Neutrinos change flavor as they journey from the Sun to the Earth through a combination of vacuum oscillations and matter-induced transformations, described by the Large Mixing Angle (LMA)-MSW solution~\cite{Wolfenstein:1977ue,Mikheyev:1985zog}. For solar neutrino data, this process is well-approximated by a two-flavor transformation model between electron neutrinos and a second component which is combination of mu/tau neutrinos~\cite{Haxton:2012wfz,Antonelli:2012qu}. 

Though the LMA-MSW model is effective at describing the combination of experimental data sets~\cite{Gann:2021ndb}, as measurements continually improve and next generation detectors come online~\cite{JUNO:2023zty,Capozzi:2018dat}, a more precise theoretical description is required. For example, the detection channels used are unable to distinguish between these $\nu_\mu$ and $\nu_\tau$ flavors. This is unfortunate, since identifying a $\nu_\tau$ component from the Sun would be particularly interesting since this mixing is one of the only mechanisms to produce $\nu_\tau$ at the MeV scale. 

Going beyond solar neutrinos, measuring the neutral current elastic scattering of $\nu_\mu$ and $\nu_\tau$ has important, broader implications. At energies characteristic of solar neutrinos, the $\nu_\mu$-electron elastic scattering cross section has been measured at terrestrial experiments~\cite{Hasert:1973cr,Ahrens:1983py,Abe:1989qk,Ahrens:1990fp,Formaggio:2012cpf}, though there has been no direct measurement of the tau neutrino scattering cross section. A measurement of this tau neutrino flux is important not only to confirm the standard model of neutrino oscillations from the Sun, but also to probe for possible effects from new physics. For example, non-standard neutrino interactions (NSI) may be important in scattering at these energies~\cite{Dutta:2020che}. 

Recent analyses have discussed the prospects of separating the $\nu_\mu$ and $\nu_\tau$ flavor components from the Sun, relying on detecting the presence of radiative corrections to the cross section through future high precision measurements of the solar neutrino flux. Ref.~\cite{Mishra:2023jlq} considered the prospects using CE$\nu$NS interactions at future large scale dark matter detectors, while Ref.~\cite{Brdar:2023ttb} considered the prospects for elastic scattering at future large scale detectors such as DUNE, Hyper-K, or JUNO. Both of these studies focused on the high-energy, $^8$B components of the solar neutrino flux, which is the most accessible component given the thresholds of the experiments. 

Given the importance such a measurement, it is interesting to take a step back and establish the sensitivity of current and past solar neutrino experiments to $\nu_\mu$/$\nu_\tau$ components. This paper is focused on establishing this sensitivity, and exploring implications for a precise measurements of the $\nu_\mu$/$\nu_\tau$ scattering cross sections. We specifically focus on the $^7$Be component of the flux as measured by Borexino~\cite{Borexino:2017rsf}, since this component is measured with high statistics, and in a region of the parameter space well-separated from experimental backgrounds. In addition, since this is a mono-energetic flux, the neutrino survival probability is straightforward to implement and use to establish sensitivity to the scattering cross section. With similar motivation, previous studies have considered the prospects for extracting new physics in the form of non-standard neutrino interactions with Borexino~\cite{Coloma:2022umy}. 

We consider mechanisms through which the elastic scattering may differ from standard 3-flavor mixing, including models in which unitarity is violated, and models with $L_\mu-L_\tau$ symmetry~\cite{Datta:2018xty}.  We expand the concept of the flavor triangle, which has been used to study 3-flavor mixing in astrophysical neutrino sources~\cite{IceCube:2015rro,Tabrizi:2020vmo}, to 3-flavor mixing in solar neutrinos. Non-unitarity has been previously investigated in the context of solar neutrinos and Borexino~\cite{Moreno:2024pbw}. We determine the exposure at which Borexino would be sensitive to separation of $\nu_\mu$ and $\nu_\tau$ fluxes, and explore the implications for both non-unitarity and $L_\mu - L_\tau$ models.  

This paper is organized as follows; Section~\ref{sec:theory} presents the theoretical models that we consider. In Section~\ref{sec:data} we present our analysis of the Borexino data, and is Section~\ref{sec:Results} we present the results of our analysis. In Section~\ref{sec:discussion} we end with discussion and implications of our results.

\section{Predicted Borexino Event-rate Observations} 
\label{sec:theory}
In this section we discuss the necessary pieces for event-rate calculations of solar neutrinos scattering elastically off electrons in the Borexino detector -- specifically we discuss interaction cross sections in the SM and beyond in~\cref{sec:CrossSections}, and calculations of the expected flavor composition of the neutrino flux at earth in~\cref{sec:Flux}.

\subsection{(B)SM Neutrino-electron Cross Sections}\label{sec:CrossSections}
We are primarily interested in the elastic scattering of SM neutrinos with electrons (EES) in the Borexino detector via the exchange of $Z$ (all flavors) and $W$ (only $\nu_e$) bosons. We will also consider modifications to this process due to a new, light mediator.

\paragraph{Tree-level EES}
At tree level the cross-section for electron scattering with $\nu_e$ is different from that with $\nu_\mu$ and $\nu_\tau$, due to the presence of charged current interaction for electron flavor in addition to the neutral current interaction, the latter of which is present for all the three flavors. At tree level the differential cross-section is given by
\begin{equation}
    \frac{d \sigma_{e\alpha}}{dT} 
    =\frac{m_e}{4\pi}\left[\left(c^{\alpha}_L\right)^2 + \left(c_R\right)^2\left(1-\frac{T}{E_\nu}\right)^2 - \left(c^{\alpha}_L\right)\left(c_R\right)\frac{m_e T}{E_\nu^2}\right]
    \label{tree level cross-section}
\end{equation}
where $T$ and $E_\nu$ are the electron recoil energy and neutrino energy respectively while $\alpha = e,\ \mu,\ \tau$ refers to the neutrino flavor. The constants $c_L^{\alpha}$ and $c_R$ are
\begin{align}
    c_L^\alpha = 2\sqrt{2}G_F\left(\sin^2\theta_w - \frac{1}{2} + \delta_{e\alpha}\right),\ \quad c_R = 2\sqrt{2}G_F \sin^2\theta_w.
\end{align}
The remaining constants are $\theta_w$ (the weak mixing angle) and $G_F$ (the Fermi constant). The Kronecker delta function $\delta_{e\alpha}$ appearing in $c_L^\alpha$ is due to the charged-current exchange available for $\nu_e$ EES and causes the overall $\nu_e$ EES cross section to be significantly larger (by nearly a factor of five) than that of $\nu_{\mu,\tau}$.

\paragraph{Radiative corrections to EES}
As seen above, for $\alpha = \mu$ or $\tau$, the differential cross section is identical. However at one-loop order (e.g., via the exchange of a $W$ loop causing a charged lepton to scatter off the target electron via the exchange of a photon), leptonic mass effects exist causing $d\sigma_{e\mu}/dT \neq d\sigma_{e\tau}/dT$. The differential cross sections may be expressed as~\cite{Tomalak:2019ibg}
\begin{equation}
    \frac{d \sigma^{\rm rad.}_{e\alpha}}{dT} = \left[1 + \frac{\alpha}{\pi}\left(\delta_v + \delta_s + \delta_I + \delta_{II}\right)\right]\frac{d \sigma^{\rm LO}_{e\alpha}}{dT} + \frac{d \sigma_{v}}{dT}  + \frac{d \sigma^{\rm dyn.}_{e\alpha}}{dT} + \frac{d \sigma^{\rm NF}}{dT}.
    \label{rad correction cross-section}
\end{equation}
Here, $d\sigma_{e\alpha}^{\rm LO}/dT$ is given by~\cref{tree level cross-section} with appropriate redefinition of $c_{L,R}$~\cite{Tomalak:2019ibg}. The quantities $\delta_{v}$ and $\delta_s$ contain information about UV-finite and gauge-independent virtual- and soft-photon corrections, respectively, and $\delta_I$ and $\delta_{II}$ provide corrections due to bremsstrahlung. The additional terms arise from QED vertex corrections ($d\sigma_v$), non-factorizable contributions to the electron energy spectrum ($d\sigma^{\rm NF}$), and from dynamical corrections due to closed fermion loops ($d\sigma_{e\alpha}^{\rm dyn.}$) The exact expressions for the above-mentioned terms can be found in Ref~\cite{Tomalak:2019ibg}. 

\begin{figure}
    \centering
    \includegraphics[width = 0.75\textwidth]{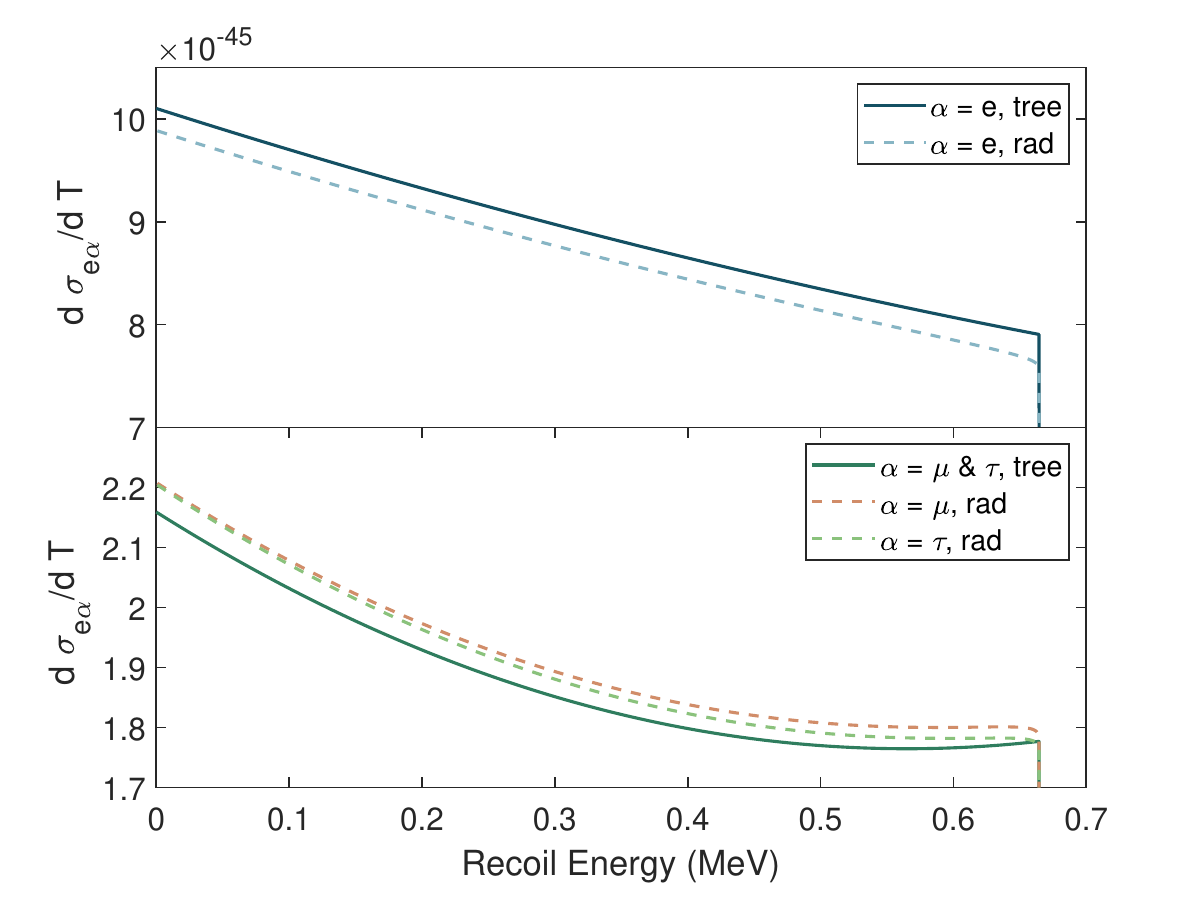}
    \caption{Differential cross-sections for neutrino-electron elastic scattering for $\nu_e$ (top) and $\nu_{\mu,\tau}$ (bottom), where the difference in panels is only the y-axis range. In each panel, we compare tree-level differential cross sections (solid lines, identical for $\nu_\mu$ and $\nu_\tau$) with those accounting for radiative corrections (dashed).\label{fig:Be7 cross-section}}
\end{figure}

Overall, the nontrivial flavor dependence (especially that which allows for differentiation between $\nu_\mu$ and $\nu_\tau$ EES) is largely present in $d\sigma_{e\alpha}^{\rm dyn.}$. In~\cref{fig:Be7 cross-section} we demonstrate these impacts for $^{7}$Be neutrinos with $E_\nu = 0.862$ MeV as a function of the recoiling electron's kinetic energy. The two panels demonstrate the same information, with the only difference being the y-axis range due to the relative strength of $\nu_e$ EES (top panel) over that of $\nu_{\mu}$ and $\nu_\tau$ (bottom panel). For each $\alpha = e,\ \mu,\ \tau$, we demonstrate the differential cross section at tree level -- solid lines,~\cref{tree level cross-section} -- and including radiative corrections -- dashed,~\cref{rad correction cross-section}. Notably, including radiative corrections ends up decreasing the $\nu_e$ EES cross section and increasing that of $\nu_\mu$ and $\nu_\tau$~\cite{Tomalak:2019ibg}, while introducing some different $T$-dependence in comparing $\nu_\mu$ (dashed orange) and $\nu_\tau$ (dashed green) rates. With precise enough measurements of the recoil energy distribution of electrons, one can, \textit{in principle}, differentiate between $\nu_\mu$ and $\nu_\tau$ EES due to these effects. Previous results have also studied these radiative corrections and their impact on coherent scattering of neutrinos, e.g., Refs.~\cite{Tomalak:2020zfh,Brdar:2023ttb}.

\paragraph{BSM Scattering with a light mediator}
Neutrino-electron scattering can be modified, especially for low momentum transfers, in the presence of new mediators that couple to both neutrinos and to electrons -- a feature present in many $U(1)$ extensions of the SM. One such extension of theoretical interest (e.g.\ due to its potential for resolving the $g-2$ anomaly of the muon) is $U(1)_{L_\mu - L_\tau}$, where the difference between muon-lepton-number and tau-lepton-number is gauged and the new boson associated with this, $Z^\prime$, gains a mass in the ${\sim}$MeV-GeV regime. This mediator will couple (at tree level) to $\nu_\mu$ and $\nu_\tau$ (but not $\nu_e$), and via loop-level effects~\cite{Kamada:2015era,Bauer:2018onh,Escudero:2019gzq}, to electrons with an induced kinetic mixing (we assume that the bare Lagrangian-level kinetic mixing between the $Z^\prime$ and the SM is zero). When we consider this scenario, we will only consider the modification of the neutrino-electron scattering in the detector and not the potential modification of matter effects for neutrinos emerging from the Sun.

In the presence of this new mediator with mass $m_{Z^\prime}$ and gauge coupling $g_{\mu\tau}$, the coefficients in~\cref{tree level cross-section} for $\nu_{\mu,\tau}$ scattering are modified~\cite{Altmannshofer:2019zhy,Amaral:2020tga}
\begin{align}\label{eq:LmuTaucLcR}
    c_{L,R} \to c_{L,R} \mp \frac{G_F}{\sqrt{2}} \frac{e^2 g_{\mu\tau}^2}{6\pi^2} \log{\left(\frac{m_\tau^2}{m_\mu^2}\right)} \frac{v^2}{m_{Z^\prime}^2 + 2m_e T},
\end{align}
where $v = 246$ GeV is the Higgs vev, $e$ is the electric charge, and the $-$ ($+$) sign corresponds to the modification for $\nu_\mu$ ($\nu_\tau$), where this new interaction provides constructive (destructive) interference with the SM. The impact on the $\nu_e e \rightarrow \nu_e e$ cross-section is negligible and can be taken to be the same as the tree-level value.

\subsection{Solar fluxes and oscillation probability}\label{sec:Flux}
The sun produces electron neutrinos via nuclear fusion processes within its core, predominantly in the ``pp reaction'' in which a proton (in the vicinity of another proton) beta decays. In the subsequent reactions, a relatively smaller fraction of electron neutrinos -- the ``$^{7}$Be neutrinos'' -- are emitted as $^{7}$Be nuclei are transformed to $^{7}$Li ones. These neutrinos are monoenergetic, $E_\nu = 0.862$ MeV, and their flux may be predicted given a solar model. In the high-metallicity model of the sun~\cite{Haxton:2012wfz}, this prediction is $5 \times 10^9$ cm$^{-2}$s$^{-1}$, with a theoretical uncertainty of $6\%$. Other decays in the sun produce large neutrino fluxes, e.g. $^{8}$B, however we focus in this work on $^{7}$Be due to its high flux and the fact that it produces monoenergetic neutrinos.

Due to their interactions with electrons in the Sun, the $\nu_e$ undergo flavor-dependent oscillations into $\nu_{\mu,\tau}$ as they travel from the core to the surface~\cite{Wolfenstein:1977ue,Mikheyev:1985zog}. This oscillation probability (due to the ``slow" change of density from core to surface) may be expressed as
\begin{align}
    P(\nu_e\rightarrow\nu_\alpha) = \sum_{i=1}^{3} P\left(\nu_e \to \nu_i\right) \left\lvert U_{\alpha i}\right\rvert^2,
    \label{eq: Prob sun}
\end{align}
where $P(\nu_e \rightarrow \nu_i)$ ($i = 1,\ 2,\ 3$) represents the adiabatic transition probability of a (flavor-eigenstate) $\nu_e$ to a (mass-eigenstate) $\nu_i$ during its propagation out of the Sun, and $\left\lvert U_{\alpha i}\right\rvert^2$ represents the overlap between the mass-eigenstate $\nu_i$ and the flavor-eigenstate $\nu_\alpha$ when the neutrino interacts in the detector. The relation between the mass eigenstates and the flavor eigenstates is  
\begin{align}
    |\nu_{i}\rangle = \sum_\alpha U_{\alpha i}|\nu_{\alpha}\rangle. 
\label{eq: nu_i U nu_fl}
\end{align}
The Hamiltonian that describes propagation in the Sun is made up of local eigenvalues and can be written as
\begin{align}
     H_{\rm fl.} = {U}E_{\rm M}{U}^\dagger + V
     \label{eq:Scrondingers eq sun}
\end{align}
The energy eigenvalue matrix (diagonal in the mass basis) is $E_{\rm M} = \frac{1}{2E_\nu} \textrm{diag}(m_1^2,m_2^2,m_3^2)$. $V$ refers to the (flavor-basis) matter potential, which is a function of electron number density and nucleon number density in the matter, and is given by 
\begin{align}
   V = \mathrm{diag}(V_{\rm cc}+V_{\rm nc},V_{\rm nc},V_{\rm nc}),\quad V_{\rm cc} = \sqrt{2} G_F n_e,\quad V_{\rm nc} = \frac{\sqrt{2}}{2} G_F n_n
    \label{eq:Vcc}
\end{align}
Consequently the $P(\nu_e\rightarrow\nu_i) $ can be written as~\cite{Blennow_2004,Mishra:2023jlq}
\begin{align}
    P(\nu_e\rightarrow\nu_i) &= \int_0^{R_\odot}\left\lvert\langle \nu_i|\nu_e(t)\rangle\right\rvert^2f(r)dr =\int_0^{R_\odot}\sum_{j=1}^3|\hat{U}_{ej}(\theta_m)|^2 P^{\rm jump}_{ji}f(r)dr
    \label{eq: Pei}
\end{align}
where $ P^{\rm jump}_{ji}$ refers to the probability of jumping from one energy eigenvector to the other. For adiabatic propagation, $P^{\rm jump}_{ji}= \delta_{ji}$. The probability is averaged over the different production zones within the Sun, weighted by the fraction of neutrinos produced in each zone $f(r)$. The information on zonal neutrino production fraction and flux information is taken from Ref.~\cite{Bahcall_rep}, and $\hat{U}_{ej}(\theta_m)$ is the eigenvector of~\cref{eq:Scrondingers eq sun}.

In the following sections, we will be analyzing the Borexino experiment and its measurement of $^{7}$Be neutrinos. We will generically be agnostic regarding the relative rate of $\nu_e$, $\nu_\mu$, and $\nu_\tau$ interactions comprising the neutrino-electron scattering for this fixed-energy portion of the analysis. However, we will be interested in comparing the extracted rates against those predicted by the flavor evolution described above. We will do so assuming that the mixing of neutrinos is unitary or considering the case where we relax that assumption. The following subsections explore those two situations in turn.

\paragraph{Oscillations with Unitary Mixing}
Assuming only three neutrinos exist, we can use the PDG parameterization~\cite{ParticleDataGroup:2022pth} to describe the $3\times 3$ unitary mixing matrix $U_{\alpha i}$ -- we assume that the three mixing angles and one CP-violating phase are determined to the level summarized by the most recent NuFit results~\cite{nu_fit,nu_fit_web}.

The survival probability of electron neutrinos from the Sun depends predominantly on the $U_{ei}$ elements, or, in terms of mixing angles, $\theta_{12}$ and $\theta_{13}$. However, the appearance probabilities into $\nu_{\mu,\tau}$ exhibit dependence on $\theta_{23}$ and $\cos\delta_{\rm CP}$. Since these two are less well-measured, there is more uncertainty on the expected fluxes of $\nu_\mu$ and $\nu_\tau$ arriving at the Earth.

If we consider the best-fit mixing angle values~\cite{nu_fit} as well as assuming three-flavor neutrino evolution through the Sun~\cite{Mishra:2023jlq,Bahcall_rep}, we arrive at the expectation for $^{7}$Be neutrinos that
\begin{align}
    P(\nu_e \to \nu_e)^{\rm BF} &= 0.528, \label{eq:PeeBF}\\
    P(\nu_e \to \nu_\mu)^{\rm BF} &= 0.204, \label{eq:PemBF}\\
    P(\nu_e \to \nu_\tau)^{\rm BF} &= 0.267. \label{eq:PetBF}
\end{align}
Given the current uncertainty on oscillation parameters~\cite{nu_fit_web}, we determine that the predicted $P(\nu_e \to \nu_e)$ has a $5\%$ relative uncertainty. We will return to the range of allowed $P(\nu_e \to \nu_{\mu,\tau})$ in later discussion.

\paragraph{Non-Unitary Mixing} \label{subsec : Non-unitary}
Considering the possible existence of one or more heavier ``sterile" neutrino states~\cite{Blennow:2016jkn,Escrihuela:2016ube}, the unitary mixing matrix may expand to an 
$n\times n$ matrix ($n$ being the total number of neutrino species). This introduces non-unitary effects when examining the mixing of the three active neutrino species, leading to a non-unitary nature within the $3\times 3$ framework.

If the mixing among the three light neutrinos is not unitary~\cite{Antusch:2006vwa,Parke:2015goa,Ellis:2020hus}, then the flavor evolution of neutrinos through the Sun, as well as the probability of interacting as a certain neutrino flavor eigenstate in an Earth-based detector, are modified. This could generically lead to an overall reduction in the expected rate of solar neutrino events at Borexino as well as in deviations from the expected relative flavor compositions in this event rate. Generally, we may parameterize the nonunitary analogue of the mixing matrix, called $N$, as
\begin{align}
    N &= \alpha U = \begin{bmatrix}
       \alpha_{11} & 0 & 0 \\
       \alpha_{21} & \alpha_{22} & 0 \\
       \alpha_{31} & \alpha_{32} & \alpha_{33}
        \end{bmatrix}  U^{3\times3}
        \label{eq: non-unitary matrix}
\end{align}
where $U^{3\times3}$ is the standard, unitary leptonic mixing matrix and unitarity is restored if $\alpha_{ij} \to \delta_{ij}$. The $\alpha_{ij}$ are constrained by numerous observations to be close to $\delta_{ij}$, but the specific value of these constraints depends on the assumed source of unitarity violation. This includes, for instance, whether a heavy neutrino inducing this violation exists below or above the electroweak scale -- we refer the interested reader to Refs.~\cite{Escrihuela:2016ube,Blennow:2016jkn} for in-depth discussion on these effects. When making predictions of the expected $^{7}$Be event rate due to non-unitarity, we will adopt the ``neutrino-only'' constraints from~\cite{Escrihuela:2016ube} which allow for ${\sim}$percent level deviations at the $3\sigma$ CL (with the exception of $\alpha_{33}$ which may be as small as approximately $0.76$ and $|\alpha_{31}|$ which may be as large as approximately $0.10$).

Importantly, any non-unitarity impacts the Hamiltonian of flavor-evolution through the Sun:~\cref{eq:Scrondingers eq sun} becomes
\begin{align*}
     H_{\rm fl.} \longrightarrow {N}E_{\rm M}{N}^\dagger + (N{N}^\dagger) V (N{N}^\dagger)
\end{align*}
This then impacts~\cref{eq: Pei} -- the eigenvectors of this Hamiltonian enter into the expected $\nu_i$ production. The final effect is in modifying the probability that a $\nu_i$ interacts ``as'' a $\nu_\alpha$ in the detector -- in the unitary scenario this probability is $\left\lvert U_{\alpha i}\right\rvert^2$, as in~\cref{eq: Prob sun}. Without unitarity, this overlap is in principle different and is given by $\left \lvert N_{\alpha i}\right\rvert^2$.

\begin{figure}
    \centering
    \includegraphics[width=0.7\linewidth]{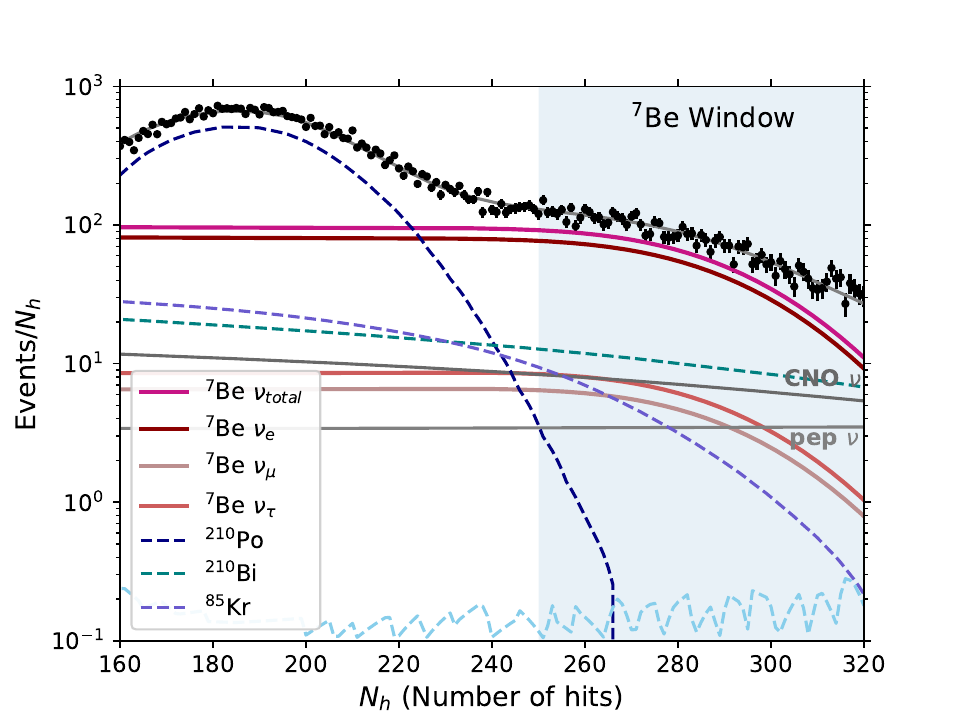}
    \caption{Expected event rate in Borexino for neutrino-electron scattering as a function of $N_h$, the number of hits. Data points, with error bars, are shown as black points. The highlighted blue region, labeled the ``$^{7}$Be Window,'' is where we analyze the data. Several flux components are highlighted, including the three different flavor contributions (as well as the total) of $^{7}$Be neutrinos, as well as different background and other solar neutrino components.}
    \label{fig:Be7 window spectrum}
\end{figure}

\section{Borexino Data Analysis}
\label{sec:data} 
In this section we describe our analysis of the Borexino data, with a particular emphasis on the $^{7}$Be solar neutrino component. Borexino is a 278-ton ultra-pure organic liquid scintillator located at LNGS in Italy. In its most recent Phase-III data set, Borexino performs a multi-component spectral analysis of the solar neutrino flux scattering off electrons in the detector, obtaining flux measurements of the $pp$, $pep$, CNO, $^{8}$B, and $^{7}$Be solar neutrino components. The $^{7}$Be flux is the best-measured, in part because it occupies a region of relatively low background with respect to the recoiling electron energy. Borexino Phase-II data have constrained the $^{7}$Be flux at the $2.7\%$ level, in comparison with, for instance,  of $pp$ neutrinos ($7.9\%$) and $^{8}$B ones ($8\%$).

Because of the low backgrounds and the mono-energetic flux, $^7$Be neutrinos allow us to precisely discern the flavor composition of the neutrino signal by removing the dependence on neutrino energy. In order to isolate the $^7$Be signal, we focus on the energy window from 250 $N_h$ to 320 $N_h$, where $N_h$ corresponds to the number of hits on the detector photomultipliers, and serves as an estimator of the recoil energy of the electron scattered by the incoming neutrino~\cite{Coloma:2022umy}. By confining our analysis to this specific range, we minimize the contributions of various backgrounds as well as neutrinos from other solar-neutrino production mechanisms (CNO, $pep$, $pp$, etc.).

Since $^7$Be neutrinos are mono-energetic($E_0$), the number of events for any flavor $\alpha$ can be written as:
\begin{equation}\label{eq:EventRate}
    N_{\alpha i} = \xi\int_{N_h^i}^{N_h^{i+1}}\int \int_{E_\nu=E_0}  \frac{d \phi_{e\alpha}(E_\nu)}{d E_\nu} \frac{d \sigma_{e\alpha}(E_\nu,T)}{d T} \frac{d \mathcal{R}(T,N_h)}{d T}d E_\nu d T d N_h = \phi P_{e\alpha} \sigma_{e\alpha}
\end{equation}
here $\xi$ is the exposure for Borexino Phase III (running for 1431.6 days, 71.3 ton fiducial volume). This quantity also contains information for detector efficiency. The term $\frac{d \mathcal{R}(T,N_h)}{d T}$ is the resolution function and $\sigma$ is the EES cross section taken from~\cref{sec:theory} -- we refer the reader to Refs.~\cite{Gonzalez_Garcia_2024,Coloma:2022umy} for more details on the resolution function, $\xi$ and $N_h(T)$. Unless otherwise stated, we include radiative corrections when considering the EES cross section.
For $^7$Be, $d\phi_{e\alpha}/dE_\nu$ is a delta function and so this term integrates trivially, giving $N_{\alpha i} \propto \phi P_{e\alpha}$.~\cref{fig:Be7 window spectrum} presents our reproduction of the Borexino expected/observed event rate with different contributions labeled. We have also separately identified the contributions to the $^{7}$Be event rate by neutrino flavor, assuming the best-fit probabilities discussed above.

We construct a $\chi^2$ test statistic that incorporates the experimental data $N_{i}^{\rm obs.}$ in the $i$th bin, as well as the total expected background ($B_i$) and signal ($S_i$) rates in that bin. For simplicity, we assume that our signal consists only of the $^{7}$Be contributions and that the background rates include the following: $^{210}$Bi, $^{85}$Kr, $^{210}$Po, Borexino's ``external'' backgrounds, as well as the solar-neutrino contributions from CNO and pep neutrinos. We also include the bin-by-bin uncertainty from Borexino, $\sigma_i$, giving our test statistic
\begin{align}
    \chi^2 = \sum_i \frac{\left(N_{i}^{\rm obs.} - \left(B_i + S_i\right)\right)^2}{\sigma_i^2}  + \chi^2_{\rm ext.},
\end{align}
where $\chi^2_{\rm ext.}$ allows for external information to be included in the form of a prior.

Since we are interested in the sensitivity of Borexino to the individual flavor components of the $^{7}$Be neutrinos, we parameterize $S_i$ as
\begin{align}\label{eq:SignalParameterization}
    S_i \equiv n \sum_{\alpha = e,\mu\tau} S_{\alpha i},\quad  S_{\alpha i}\equiv \frac{f_\alpha}{P_{e\alpha}^{\rm BF}} N_{\alpha i}^{\rm BF},
\end{align}
where $P_{e\alpha}^{\rm BF}$ are the best-fit expected oscillation probabilities cf~\cref{eq:PeeBF,eq:PemBF,eq:PetBF} and $N_{\alpha i}^{\rm BF}$ is the expected event rate in bin $i$ for flavor $\alpha$ assuming the best-fit oscillation probability in~\cref{eq:EventRate}. The overall normalization parameter $n$ that rescales all of the $^{7}$Be components is to account for uncertainty in the overall flux of the neutrinos from the Sun. Generically, this allows for four free parameters -- $\left\lbrace n,\ f_e,\ f_\mu,\ f_\tau\right\rbrace$ -- however, considering all four to vary simultaneously will result in perfect degeneracies among them making it impossible to constrain each independently. In the following subsection, we will detail the choices we make regarding which parameter(s) to consider and what external information to include in $\chi^2_{\rm ext.}$. We will also discuss how our measurements of these four parameters allow for direct comparisons of expected, allowed values of the oscillation probabilities $P_{e\alpha}$.

In general, we will perform our analyses keeping the background rates $B_i$ fixed. In all of the cases that we present, we have also considered allowing one background at a time to vary with a Gaussian prior on its normalization at the $10\%$ level. We find that this does not significantly impact the results of our analysis and so we opt to keep the normalizations fixed in what we display.

\section{Analysis Choices \& Results}\label{sec:Results}
In this section we detail the analysis choices we make when studying Borexino and its sensitivity to the $^{7}$Be flux. We perform a number of different analyses, beginning with a simplified two-flavor approach, followed by two distinct three-flavor studies. Finally, we explore the Borexino sensitivity to new gauge bosons ($L_{\mu} - L_{\tau}$) and the impact of other underlying model parameters to this sensitivity.

\subsection{Two-flavor Analysis}\label{subsec:TwoFlavor}
We first consider an effective two-flavor analysis in which we divide the signal into electron-flavor contributions ($\nu_e$) and ``others'' ($\nu_x$). Instead of considering $S_{\mu i}$ and $S_{\tau i}$ separately as in~\cref{eq:SignalParameterization}, we define an effective $S_{xi}$ using
\begin{align}
    S_{xi} \equiv \frac{f_{x}}{P_{e\mu}^{\rm BF} + P_{e\tau}^{\rm BF}} \left(N_{\mu i}^{\rm BF} + N_{\tau i}^{\rm BF}\right)
\end{align}
If we fix the normalization such that $n = 1$, then the $f_{\alpha}$ parameters map on directly to the oscillation probabilities $P_{e\alpha}$. In that case, we have a number of theoretical expectations against which we can compare. First, unitary mixing requires $f_e + f_x = 1$. If we consider knowledge of three-neutrino mixing and constraints from other solar neutrino experiments, then there is a further restriction on the range of the allowed $f_e$ within this range. However, if we allow for non-unitary mixing, more of the $f_e$ vs. $f_x$ parameter space is allowed. We will consider two scenarios with $n=1$ fixed -- one in which $f_e$ is allowed to vary freely and one in which it is constrained, adding a prior $\chi^2_{\rm ext.} = (f_e - P_{ee}^{\rm BF.})^2/\sigma_{e}^2$, with $\sigma_e = P_{ee}^{\rm BF.}\times 5\%$.
\begin{figure}[!htbp]
    \includegraphics[width=0.48\linewidth]{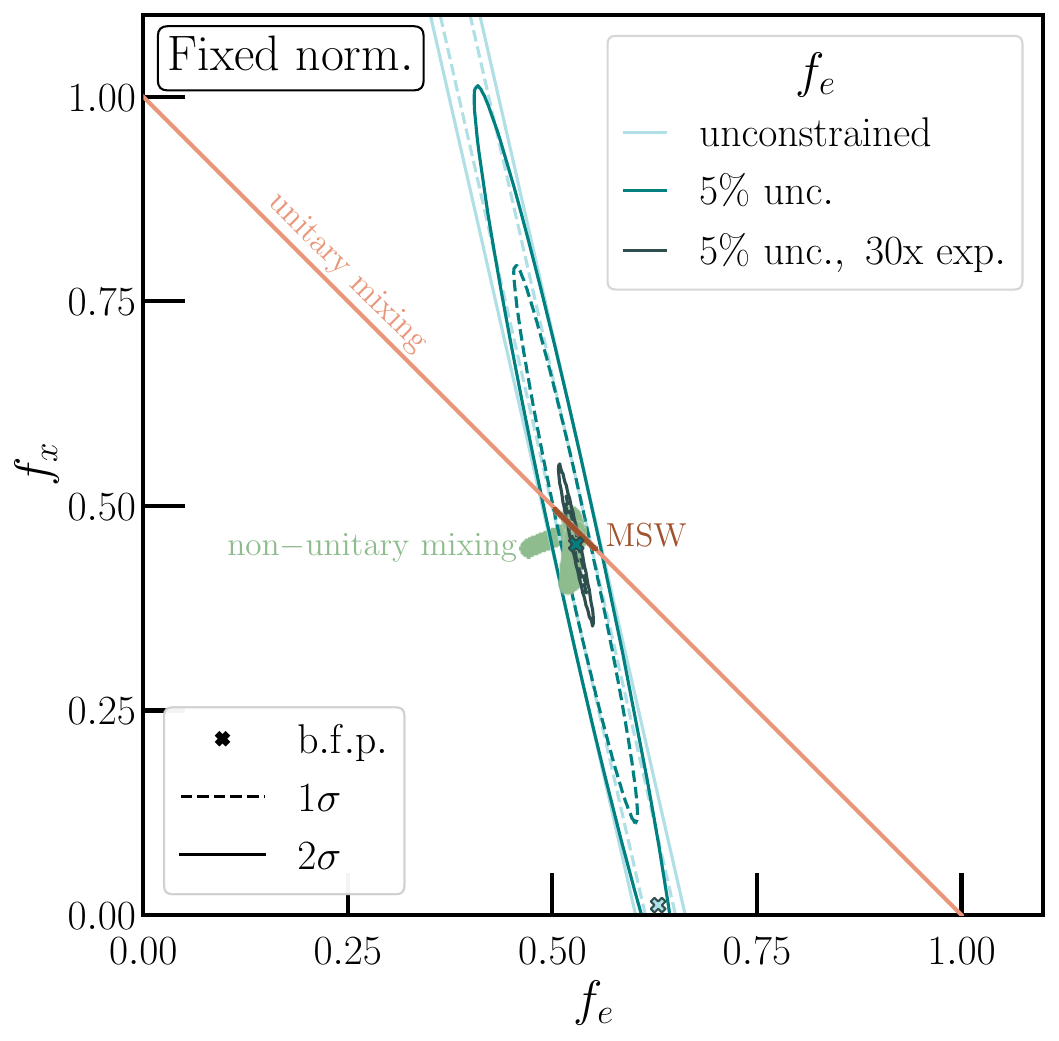}
    \includegraphics[width=0.48\linewidth]{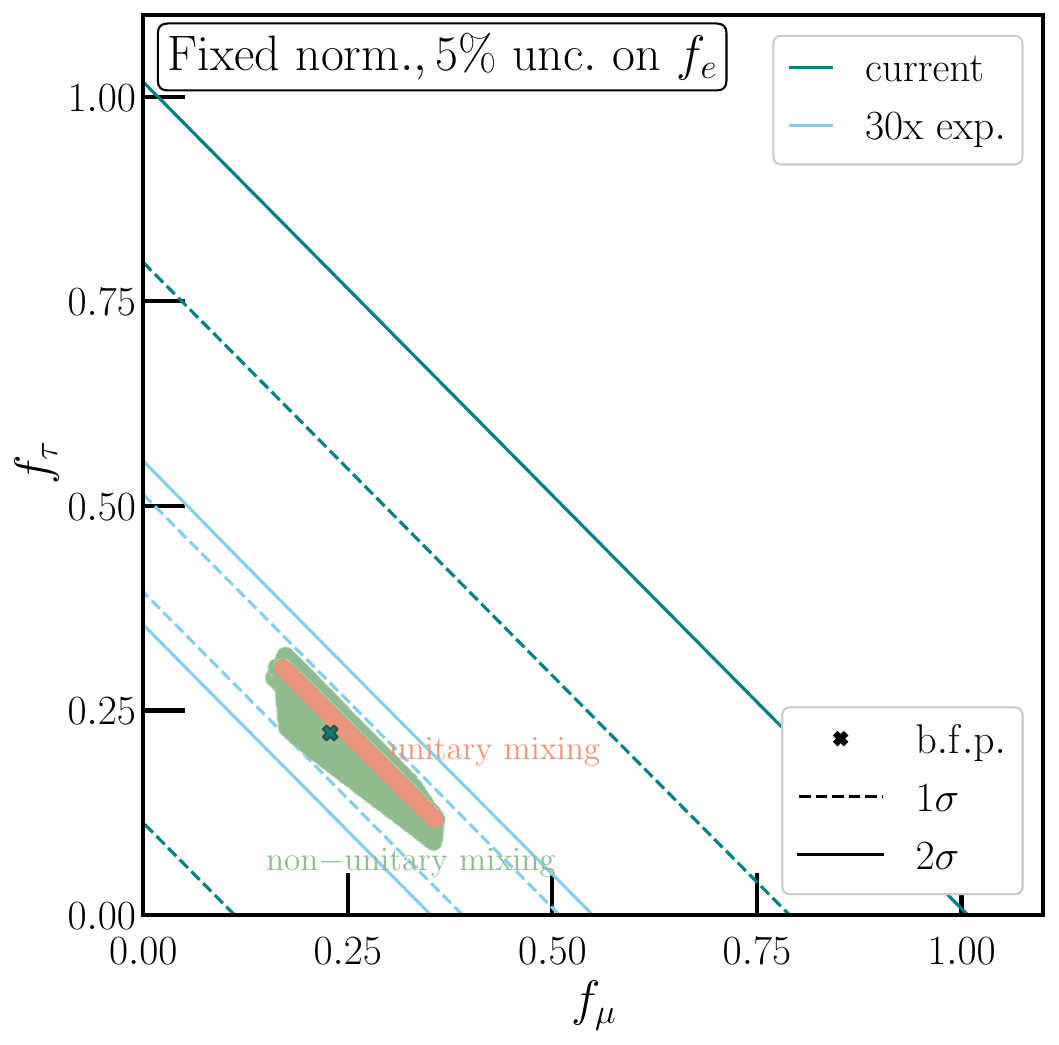}
    \caption{Measurements of Borexino Phase-III at $1\sigma$ (dashed) and $2\sigma$ (solid) confidence for a number of analysis scenarios. Left panel: the two-flavor analysis of~\cref{subsec:TwoFlavor}. We consider three different scenarios, all with the normalization $n$ fixed -- one in which $f_e$ is fully unconstrained (light blue), one in which a $5\%$ prior is included (dark blue), and one in which a further $30$x factor is included on the exposure (dark gray). Best-fit points are marked with an 'x.' In the left panel, we compare with expected regions of parameter space when considering unitarity (orange), the MSW solution (brown), and allowing for present-day uncertainty of non-unitarity (green). Right panel: the three-flavor analysis of~\cref{subsec:ThreeFlavor}, where $n$ is fixed and $f_e$ is profiled subject to a $5\%$ uncertainty. The dark blue (light blue) contours display measurements with current ($30$x exposure) data, compared against predicted points in parameter space for unitary (orange) or non-unitary (green) mixing. See text for further detail.}
    \label{fig:fefx_fmft}
\end{figure}

Our results are shown in the left panel of~\cref{fig:fefx_fmft} for the case where $f_e$ is left unconstrained (light blue) vs. when a $5\%$ prior is included (dark blue). For both cases, we show the best-fit point (as an `x') along with $1\sigma$ (dashed) and $2\sigma$ (solid) CL contours. We also display a hypothetical measurement that could be performed with $30$ times the exposure of Borexino (dark gray), assuming that uncertainties are dominantly statistical so that the test-statistic $\chi^2$ scales linearly with this factor of 30. In comparison, we show (orange) the line in parameter space for which these measurements are consistent with unitarity, $f_e + f_x = 1$, with the MSW solution for higher-energy solar neutrinos (brown), as well as points in parameter space allowed when current constraints on non-unitary mixing are considered (green). While some of these points exist outside of our current $1/2\sigma$ CL contours, we do not purport to improve on existing non-unitarity constraints, as these measurement contours can expand significantly if we allow the normalization $n$ to vary, even at the $6\%$ level.

In the left panel of~\cref{fig:fefx_fmft}, we note that, without any external information on $f_e$, the Borexino data are consistent with $f_x = 0$ (i.e., no $\nu_\mu$ or $\nu_\tau$ contribution) at $1\sigma$ CL -- this is driven both by (a) the relatively similar shape between $\nu_e$ and $\nu_{\mu,\tau}$ contributions to the Borexino event rate, as in~\cref{fig:Be7 window spectrum}, as well as (b) the fact that the $\nu_e$ EES cross section is significantly larger than that of $\nu_{\mu,\tau}$. However, once the $5\%$ prior is included (dark blue contours), we can exclude $f_x = 0$ at greater than $1\sigma$ CL and the best-fit point moves very close to the region predicted by unitarity and the MSW solution. Finally, we note that this analysis includes the presence of radiative corrections in the $\nu_\alpha$ EES cross sections. This decision does not impact the measurements of $f_e$ significantly, nor the uncertainty on the measurement of $f_x$; however, it does shift the best-fit value of $f_x$ substantially -- this is summarized in~\cref{tab:RCTable}
\begin{table}[]
    \centering
    \begin{tabular}{c||c|c}
    Parameter & Measurement without Radiative Corrections & Measurement with Radiative Corrections \\ \hline\hline
    $f_e$ & $0.529 \pm 0.0762\ (0.1245)$ & $0.529 \pm 0.0762\ (0.1246)$ \\ \hline
    $f_x$ & $0.3889 \pm 0.3545\ (0.583)$ & $0.4529 \pm 0.341\ (0.561)$ \\ \hline
    \end{tabular}
    \caption{Change in extracted measurements of $f_e$ vs. $f_x$ (in the analysis of~\cref{subsec:TwoFlavor}) when radiative corrections are or are not included in the calculation of the EES cross sections. In the measurement uncertainties, the first number (second number in parentheses) corresponds to the $1\sigma$ ($2\sigma$) CL uncertainty.}
    \label{tab:RCTable}
\end{table}

\subsection{Three-flavor Analysis}\label{subsec:ThreeFlavor}
We also consider a more complete three-flavor analysis, incorporating contributions from all three neutrino flavors: $\nu_e$, $\nu_\mu$, and $\nu_\tau$. We allow $f_{e}$, $f_{\mu}$, and $f_{\tau}$ to vary, fixing the normalization $n$. In that case, $f_{\alpha}$ correspond directly to inferred measurements of $P_{e\alpha}$. Since $P_{e\alpha}$ may be effectively determined by other solar neutrino experiments, we will include a $5\%$ prior on $f_e$ as in the above via $\chi^2_{\rm ext.}$. 

The resulting measurements of $f_\mu$ and $f_\tau$, after profiling over $f_e$, are shown in the right panel of~\cref{fig:fefx_fmft} -- we show this assuming current experimental data (dark blue) and with a hypothetical factor of 30 larger exposure as in the above (light blue). As with the above, we show $1\sigma$ (dashed) and $2\sigma$ (solid) CL contours in both cases. Unsurprisingly, we find that there is not significant power in distinguishing the $\nu_\mu$ vs. $\nu_\tau$ components given current experimental data, or even with this significantly larger exposure considered. The allowed regions of parameter space assuming unitary mixing (orange, with the variance of $\theta_{23}$ and $\delta_{\rm CP}$ driving the motion) and non-unitary mixing (green) are superimposed here -- we see that especially when considering current data, these hypotheses are very well consistent with our measurements. We also observe that the point $(f_\mu,\ f_\tau) = (0,\ 0)$ is disfavored at over $1\sigma$ CL (consistent with how in the two-flavor case, we saw $f_x = 0$ was disfavored when a $5\%$ prior on $f_e$ was included) -- here we can quantify that it is disfavored relative to the best-fit-point by $\Delta\chi^2 = 4.09$. However, if we repeat this same analysis using only tree-level calculations for the $\nu_\alpha$ EES cross sections, we see that this point is only disfavored at the level of $\Delta\chi^2 = 2.76$ -- including radiative corrections in our $\nu_\mu$ and $\nu_\tau$ EES cross sections makes these relatively small contributions easier to detect.

\subsection{Flavor Triangle Measurements}\label{subsec:FlavorTriangle}
In a similar fashion to our three-flavor analysis of~\cref{subsec:ThreeFlavor}, we can choose to scan over a different set of parameters -- $n$ as well as two of the three $f_\alpha$, assuming that the third is constrained by $\sum_\alpha f_\alpha = 1$. In doing so, we are assuming that the flux normalization can vary (subject to a $6\%$ prior given theoretical uncertainty~\cite{Haxton:2012wfz}) and that all of the events present are due to electron neutrinos, or their oscillation into $\nu_\mu$ or $\nu_\tau$. By making this choice, we may present our measurement in terms of a ternary flavor triangle, as done e.g.\ with high-energy astrophysical neutrinos with IceCube~\cite{IceCube:2015rro,Arguelles:2015dca,Bustamante:2016ciw,Song:2020nfh} or with diffuse supernova neutrinos~\cite{Tabrizi:2020vmo}. 

\begin{figure}[!htbp]
    \includegraphics[width=0.55\linewidth]{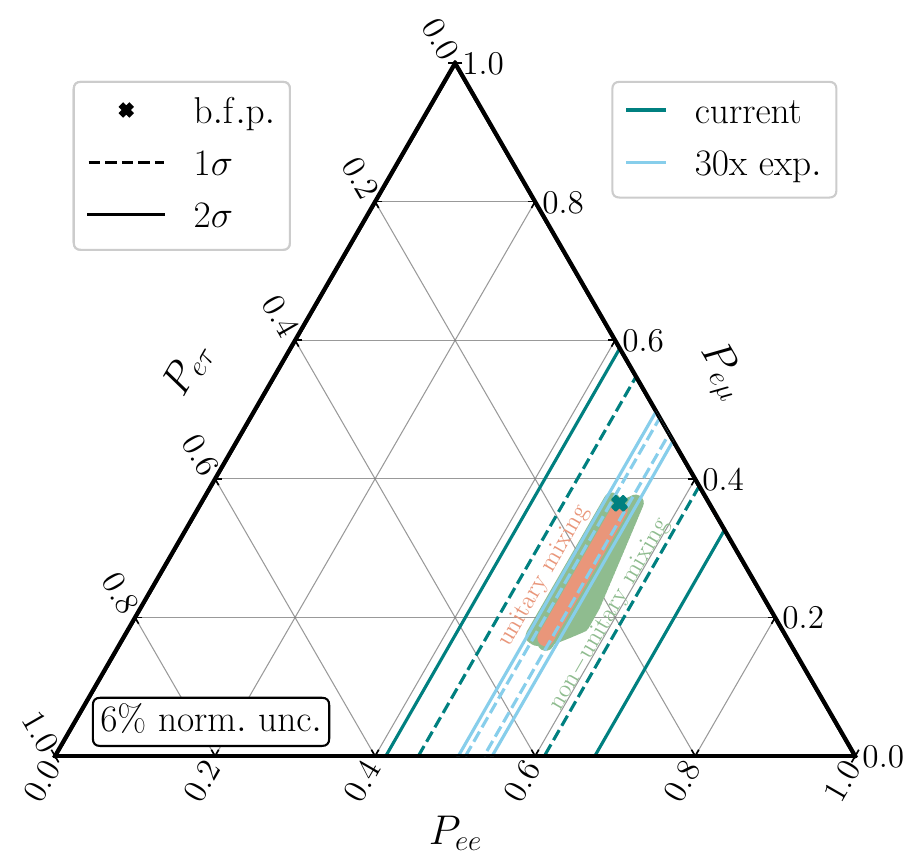}
    \caption{Flavor triangle showing the measurement capability of Borexino Phase-III at $1\sigma$ (dashed) and $2\sigma$ (solid), assuming present data (dark blue) as well as a hypothetical $30$x exposure (light blue). In both cases, a $6\%$ prior is included on the normalization $n$ which is profiled, and the sum of the probabilities $\sum_\alpha P_{e\alpha}$ is fixed to be one. Colored regions indicate the portions of parameter space expected under unitary (orange) and non-unitary (green) mixing, subject to present-day uncertainties.}
    \label{fig:ternary}
\end{figure}

The resulting measurements presented as a flavor triangle are shown in~\cref{fig:ternary}, where each point on the figure automatically guarantees $\sum_\alpha f_\alpha = 1$. Since we also allow $n$ to vary, subject to a $6\%$ Gaussian prior, we choose to interpret the $f_\alpha$ as probabilities $P_{e\alpha}$. In this figure, the dark blue contours (dashed: $1\sigma$ CL, solid: $2\sigma$ CL) correspond to our current capabilities using $^{7}$Be neutrinos at Borexino, whereas the light blue contours demonstrate the possibility of a factor of 30 larger exposure as considered in the previous two subsections. Even though the only prior information we consider is this $6\%$ uncertainty on $N$, we find that we can reasonably constrain $f_e$ to be between ${\sim}[0.4,\ 0.7]$ at $2\sigma$ CL. We contrast these measurements against the predicted relative event rates/oscillation probabilities when we consider unitary, three-neutrino mixing (orange), as well as going beyond to allow for non-unitary mixing (green). Since $n$ is allowed to vary (in contrast to the left panel of~\cref{fig:fefx_fmft} in our two-flavor analysis), we see that these non-unitary points are all within the current measurement's $1\sigma$ CL contours. However, we see that with this hypothetical factor of 30 larger exposure, we begin to probe this region more significantly.

In an ideal world, such a measurement would also allow for distinction between $P_{e\mu}$ and $P_{e\tau}$, but with the current tools we have, the data are consistent with either of these being zero -- the test-statistic is very flat along the diagonal direction connecting between $P_{e\mu} = 0$ and $P_{e\tau} = 0$. Even with a substantially larger exposure (beyond the $30$x we have presented here), this separation is very challenging. We find that with an increase in the exposure by a factor of approximately $3000$, this differentiation would be possible. Given this, we conclude that other mechanisms are likely necessary to detect the difference between $\nu_\mu$ and $\nu_\tau$ EES processes if working within the standard model, where their differences are relatively small and only separated due to radiative corrections.

\subsection{$L_\mu-L_\tau$ Analysis}\label{subsec:LmuLtau}
Finally, when considering a light gauge boson coupled to $L_\mu - L_\tau$, as in~\cref{eq:LmuTaucLcR}, we consider the same $S_{\alpha i}$ but now allowing $N_{\mu i}$ and $N_{\tau i}$ to depend on the new gauge boson's mass $m_{Z^{'}}$ and coupling $g_{\mu\tau}$. This enters via the differential cross section in~\cref{eq:EventRate}. In the most general sense, we are sensitive to five free parameters: these two new-physics ones, and three among $\left\lbrace n, f_{e}, f_{\mu}, f_{\tau}\right\rbrace$. In order to compare against previous literature results, we will consider a number of different analysis choices regarding sensitivity to $\left\lbrace m_{Z^{'}},g_{\mu\tau}\right\rbrace$. First, we will fix $\left\lbrace n,f_e, f_\mu, f_\tau\right\rbrace$ to their best-fit values -- this will serve as a direct comparison to other analyses of Borexino and $L_{\mu} - L_{\tau}$ gauge bosons, however we expect that it will yield the most aggressive constraints. Then, we will consider three different sets of prior knowledge in the form of $\chi^2_{\rm ext.}$, in all cases constraining $\sum_\alpha f_\alpha = 1$:
\begin{itemize}
    \item Include $6\%$ normalization uncertainty on $n$~\cite{Haxton:2012wfz}, allowing $f_e$ and $f_\mu$ to vary freely between $[0, 1]$.
    \item Include $5\%$ relative uncertainty on $f_e$ (from theoretical knowledge of oscillation parameters impacting $P_{ee}$) but letting $n$ vary freely between $[0, 2]$ and $f_{\mu}$ to vary freely as well.
    \item Combining the two above cases, $6\%$ uncertainty on $n$ and $5\%$ on $f_e$.
\end{itemize}

\begin{figure}
    \centering
    \includegraphics[width=0.6\linewidth]{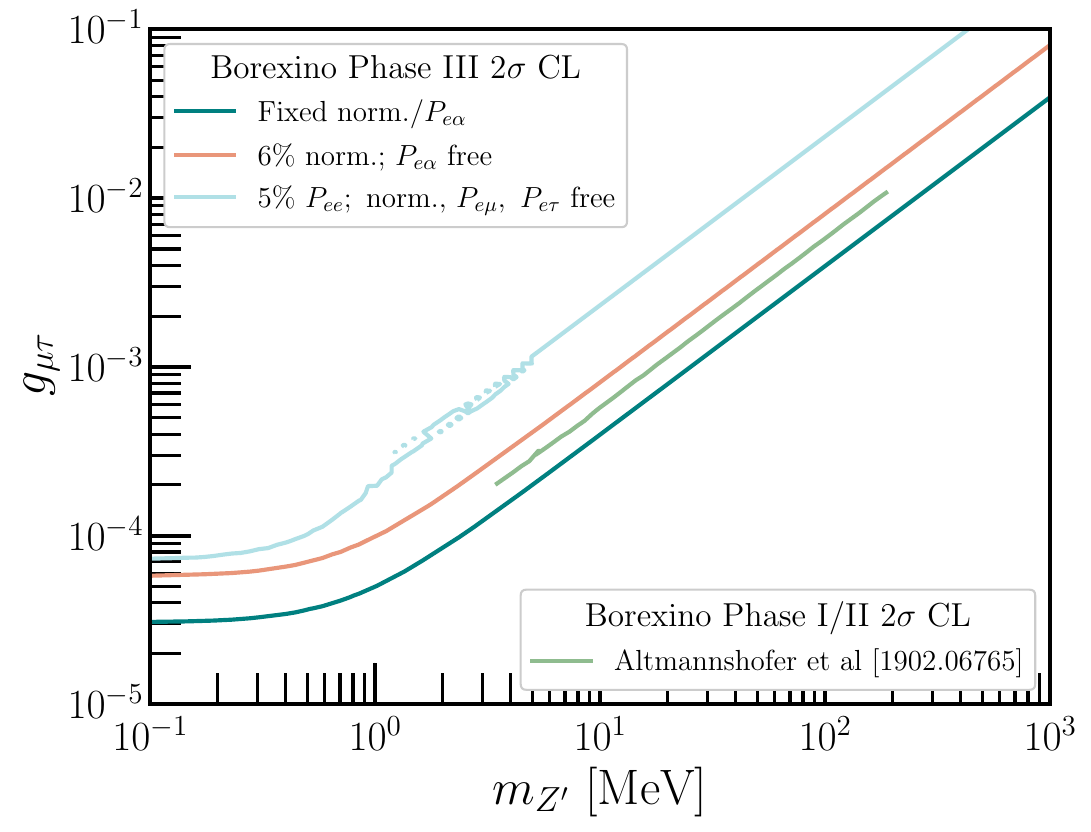}
    \caption{ Constraints on the $L_\mu-L_\tau$ model from the Borexino data. The dark blue is the $2\sigma$ contour for Borexino Phase III data using $L_\mu-L_\tau$ model, keeping the probabilities and flux normalization fixed to the best-fit value. The light blue and orange lines are $2\sigma$ contours calculated by scanning over $M_{Z'}$ and $g_{Z'}$, $n$, $p_e$ and $p_\mu$, with having a 5$\%$ Gaussian prior on $p_e$ and a 6$\%$ on $n$ respectively. The green line is the previous estimation based on Phase I and Phase II data.~\cite{Altmannshofer:2019zhy}} 
    \label{fig:Lmu Ltau }
\end{figure}
The resulting $2\sigma$ constraints we derive in these different situations are shown in~\cref{fig:Lmu Ltau }, in comparison with results analyzing previous Borexino data~\cite{Altmannshofer:2019zhy} (green). We see significant (factors of ${\sim}2$ in the size of $g_{\mu\tau}$) differences in the derived constraints depending on what prior information we include. If we fix the expected normalization as well as the oscillation probabilities (in the form of $P_{e\alpha}$), we arrive at a result (dark blue) that is consistent with the previous results of Ref.~\cite{Altmannshofer:2019zhy}, given the increase in exposure in this data release. However, if we allow the normalization $n$ to vary freely as well as the oscillation probabilities (subject to $\sum_\alpha P_{e\alpha} = 1$), we find that the Borexino data cannot constrain this model at all. This is because, without any external prior information, the data are consistent with the entirety of the $^{7}$Be events coming from $\nu_e$ EES and $P_{e\mu} = P_{e\tau} = 0$ (as we saw in~\cref{subsec:TwoFlavor}). If there are no $\nu_{\mu,\tau}$, then we cannot test the $U(1)_{L_\mu - L_\tau}$ model. However, things improve when prior information is included. For instance, if we still allow the normalization to vary freely but constrain $P_{ee}$ at the $5\%$ level, we can constrain this model as shown by the light blue line -- we consider this to be a very conservative constraint. If instead of including prior information on $P_{ee}$, we do so on the normalization $n$ at the $6\%$ level, we arrive at the orange curve, approximately a factor of $2$ weaker than the everything-fixed result in dark blue. Finally, we have also considered including priors on both $P_{ee}$ and $n$ -- that result is degenerate with the orange curve seen here. Overall we find that while Borexino is capable of using its $^{7}$Be neutrino events to constrain this new-physics scenario, some precaution is warranted given the uncertain flavor composition of these low-energy neutrinos.

\section{Discussion \& Conclusions}
\label{sec:discussion}
In this paper we have performed a full 3-flavor analysis of Borexino Phase-III data, focusing on the $^{7}$Be solar neutrino signal. We have focused on $^{7}$Be because it is a mono-energetic line and is confined to a region of relatively low backgrounds. We considered the impact of radiative corrections in the calculation of $\nu_\mu$ and $\nu_\tau$ elastic scattering with electrons, allowing us to gauge the current sensitivity of high-precision solar neutrino data to differentiate between these two processes. While the current results are not able to distinguish $\nu_\mu$ vs. $\nu_\tau$ elastic scattering, we find that a significantly larger exposure, approximately 3000 times that of Borexino, would be able to perform this exciting differentiation. We have also found that accounting for radiative corrections in our calculations makes the presence of $\nu_\mu$ and $\nu_\tau$ fluxes easier to detect than if they are ignored in such an analysis.

Throughout our work, we have considered both a simplified, two-flavor approach (combining the $\nu_\mu$ and $\nu_\tau$ components to compare them against the $\nu_e$ one) as well as considering these three contributions independently. In all of these analyses, we have compared our resulting measurements of these components against expectations, both within the standard, three-neutrino (unitary evolution) framework, as well as going beyond. We have demonstrated, for instance, that with $\mathcal{O}(30)$ times larger exposure, an experiment such as Borexino is sensitive to new levels of non-unitarity present in the leptonic sector. In constructing these analyses, we have also presented for the first time a flavor triangle in terms of the flavor components as they arrive at the Earth for low-energy, solar neutrinos coming from $^{7}$Be decays. Comparisons of these flavor triangles for other low-energy components will yield fruitful information as measurements improve in the coming era.

Finally, we have also considered the possibility of beyond-the-standard-model contributions to low-energy neutrino-electron scattering, with an emphasis on models that primarily impact $\nu_\mu$ and $\nu_\tau$ scattering. The demonstrative model that we have focused on is the $U(1)$ extension of the SM in which $L_\mu - L_\tau$ is gauged, allowing for new interactions between $\nu_{\mu,\tau}$ and electrons to arise from a light, MeV-scale mediator. We have performed comparisons against existing constraints using Borexino's $^{7}$Be neutrino sample on this model, and we demonstrate consistency, however we preach caution due to the impact of uncertainties (e.g., on the flavor composition of the $^{7}$Be neutrinos) in arriving at such constraints.

While bounding non-unitary models may not be possible for Borexino, future solar neutrino experiments should be able to improve bounds. In particular, the JUNO experiment~\cite{JUNO:2023zty} will be sensitive to $^{7}$Be solar neutrinos, and it may be possible to perform a similar analysis on the JUNO signals and backgrounds as performed in this paper. As a crude estimate, with ${\sim}10$ years of data taking, for an ideal background scenario at JUNO, the flux uncertainty will be reduced by over an order of magnitude, which would probe the non-unitarity regime discussed in our analysis. Additional possibilities exist in studying $^{8}$B neutrinos at DUNE or lower-energy $pp$ neutrinos at future Xenon-/Argon-based dark matter direct detection experiments. However, these signals do not have the advantage exploited throughout this work from the monoenergetic nature of the $^{7}$Be neutrinos. Nevertheless, future analyses of this nature as we progress into the next generation of neutrino and dark matter experiments will naturally extend our understanding of solar neutrinos and their flavor compositions.

\section*{Acknowledgements} 
The authors are all supported by the DOE Grant No. DE-SC0010813. We are very grateful to O. Tomalak and R. Plestid for several discussions on radiative corrections for EES. 

\bibliographystyle{JHEP}
\bibliography{apssamp}

\end{document}